\documentstyle[12pt]{article}
\newcommand{\eq}{\begin{equation}}
\newcommand{\en}{\end{equation}}
\newcommand{\eqn}{\begin{eqnarray}}
\newcommand{\enn}{\end{eqnarray}}
\newcommand{\CR}{\nonumber \\}
\newcommand{\bchi}{\bar{\chi}}

\newcommand{\I}{{\rm i}}
\newcommand{\half}{{1\over2}}
\newcommand{\pa}{\partial}
\newcommand{\bvp}{\bar{\varphi}}
\newcommand{\A}{\alpha}
\newcommand{\B}{\beta}
\newcommand{\D}{\delta}

\newcommand{\lm}{\lambda}
\newcommand{\vp}{\varphi}
\newcommand{\ord}[2]{{#2 \over (z-w)^{#1}}}
\newcommand{\ordo}[1]{{#1 \over z-w}}
\newcommand{\tW}{\widetilde{W}}
\newcommand{\cO}{{\cal O}}

\relax
\begin{document}
\renewcommand{\thefootnote}{\fnsymbol{footnote}}
\begin{titlepage}
\null
\begin{flushright}
NBI-HE-93-08 \\
February 1993
\end{flushright}
\vspace{3cm}
\begin{center}
{\Large
Free Field Realization of $N=2$ Super $W_{3}$ Algebra
\par}
\lineskip .75em
\vskip 3em
\normalsize
{\large Katsushi Ito}\footnote{Address after March 1, 1993:
Institute of Physics, University of Tsukuba, Ibaraki 305, Japan \\
E-mail: ito@nbivax.nbi.dk}
\vskip 3em
{\it Niels Bohr Institute, Blegdamsvej 17, DK-2100,
Copenhagen {\O}, Denmark}
\vskip 1.5em
{\bf Abstract}
\end{center} \par
We study the quantum $N=2$ super-$W_{3}$ algebra using
the free field realization, which is obtained from the
supersymmetric Miura transformation associated with the Lie
superalgebra $A(2|1)$.
We compute the full operator product expansions of the algebra
explicitly.
It is found that the results agree with those
obtained by the OPE method.

\end{titlepage}
\baselineskip=0.8cm
\renewcommand{\thefootnote}{\arabic{footnote}}
\setcounter{footnote}{0}
$W$-algebraic extensions of the $N=2$ superconformal algebra
\cite{InMaYa} have attracted much interest in the context of
conformal field theories\cite{Ito1,NeYa}, topological
$W$-gravities\cite{Li} and more recently in non-critical
$W$-strings\cite{BeLeNeWa}.
There are at least two approaches to study quantum $W$-algebras\cite{BoSc}.
One is based on the OPE method\cite{Za}, in which $W$-algebras can be
constructed by requiring the associativity and closure of the algebra.
The other approach is the free field realization\cite{FaZa}, in which
the generators are expressed by free fields
through the quantum Miura transformations.
In this approach the closure of the algebra is not obvious at
quantum level, in particular for the $W$-algebras associated with
non simply laced Lie algebra\cite{KaWa}.

In the case of $N=2$ super-$W$ algebras, it has been understood that
the free field realization can be obtained from the quantum hamiltonian
reduction \cite{BeOo} of affine Lie superalgebras $A(n|n-1)^{(1)}$
\cite{Ito1,NeYa}.
The classical Poisson brackets structure in the case of $A(2|1)^{(1)}$
has been studied by using techniques such as the super Gel'fand-Dickii
bracket \cite{InKa} and Polyakov's soldering procedure \cite{AM,Ito2}.
In the quantum case, the full algebra has been presented by using the
OPE method \cite{Ro,Odake}.
In the free field approach, however, only some of the operator product
expansions have been calculated \cite{NeYa}.

In this letter, we study the free field realization of $N=2$
super-$W_{3}$ algebra and compute the full operator product expansions.
We find that the present free field approach gives the same algebra
as the one obtained by the OPE method.

We begin with the supersymmetric Miura transformation based on
a Lie superalgebra $A(n|n-1)=sl(n+1|n)$.
We take the purely odd simple root system
$\{ \A_{1}, \ldots, \A_{2n} \}$,
with the Cartan matrix $\A_{i}\cdot\A_{j}=(-1)^{i-1}\D_{i+1,j}$.
Let $\{ \lm_{1}, \ldots, \lm_{2n} \}$ be the fundamental weights of
$A(n|n-1)$ satisfying $\A_{i}\cdot\lm_{j}=\D_{i j}$.
Denote $N=1$ super-holomorphic coordinate by $Z=(z,\theta)$
and a super-derivative by $D={\pa\over\pa\theta}+\theta\pa$
($\pa\equiv{\pa\over\pa z}$).
We introduce $2n$ free bosons
$\phi(z)=(\phi_{1}(z), \ldots, \phi_{2n}(z))$
and real fermions $\psi(z)=(\psi_{1}(z), \ldots, \psi_{2n}(z))$
satisfying the operator product expansions (OPEs):
\eq
\phi_{i}(z)\phi_{j}(w)=-\D_{i j}{\rm ln}(z-w)+\ldots, \quad
\psi_{i}(z)\psi_{j}(w)={\D_{i j}\over z-w}+\ldots \ .
\en
Define the scalar superfields by $\Phi(Z)=\phi(z)+\I \theta \psi(z)$.
The $A(n|n-1)$ super Miura transformation \cite{EvHo} is defined as
\eq
D^{2n+1}+
\sum_{i=2}^{2n+1}W_{{i\over 2}}(Z)(a D)^{2n+1-i}
= :(a D-\Theta_{2n+1}(Z))(a D-\Theta_{2n}(Z))\cdots (a D-\Theta_{1}(Z)):,
\label{eq:miura}
\en
where $\Theta_{i}(Z)=(-1)^{i-1}(\lm_{i}-\lm_{i-1})\cdot D\Phi (Z)$
($\lm_{0}=\lm_{2n+1}\equiv 0$) and $a=-\I\A_{+}$.
The symbol $:\ :$ denotes the normal ordering.
The currents $W_{i/2}(Z)$ ($i=2, \ldots, 2n+1$) generate
the $N=2$ super $W_{n+1}$ algebra.
In the following we consider the case $n=2$ and use the component
formalism.
Let us define  bosonic and fermionic $W$-currents as follows:
\eqn
W_{1}(Z)&=& J(z)+\I \theta [G^{+}(z)+G^{-}(z)], \CR
W_{{3\over 2}}(Z)&=& a [\I G^{-}(z)+\theta T(z)] , \CR
W_{2}(Z)&=& W_{2}(z)+\I\theta [W^{+}(z)+W^{-}(z)], \CR
W_{{5\over 2}}(Z)&=& a [\I W^{-}(z)+\theta W_{3}(z)].
\label{eq:wcu}
\enn
It is convenient to introduce complex bosons $\vp_{i}(z),\bvp_{i}(z)$
($i=1,2$) and complex fermions $\chi_{i}(z)$, $\bchi_{i}(z)$ ($i=1,2$)
by $\vp_{i}=\lm_{2i}\cdot\phi(z)$, $\bvp_{i}=\A_{2i}\cdot\phi(z)$,
$\chi_{i}=\lm_{2i}\cdot\psi(z)$ and  $\bchi_{i}=\A_{2i}\cdot\psi(z)$.
{}From the Miura transformation (\ref{eq:miura}), we find that
the generators of the $N=2$ super-$W_{3}$ algebra are given by
\eqn
J&=& U_{1}+U_{2}-a\pa\bvp_{2}, \CR
G^{+}&=& G^{+}_{1}+G^{+}_{2}, \CR
G^{-}&=& G^{-}_{1}+G^{-}_{2}-a\pa\bchi_{2}, \CR
T&=& T_{1}+T_{2}-a\pa^{2}\bvp_{2}, \CR
W_{2}&=& a^{2}\pa U_{1} -a\bchi_{2}G^{+}_{1}
        +U_{1}U_{2}-a^{3}\pa^{2}\bvp_{2}-a\pa\bvp_{2}U_{2}, \CR
W^{+}&=& a^{2}\pa G^{+}_{1}+a\pa\bvp_{2}G^{+}_{1}
              +U_{2}G^{+}_{1}+U_{1}G^{+}_{2}-a\pa\bvp_{2}G^{+}_{2}, \CR
W^{-}&=&a^{2}\pa G^{-}_{1}-a^{3}\pa^{2}\bchi_{2}+U_{2}G^{-}_{1}
              +U_{1}G^{-}_{2}+a\bchi_{2}T_{1}-a\bchi_{2}\pa U_{1}
              -a U_{2}\pa\bchi_{2}-a\pa\bvp_{2}G^{-}_{2}, \CR
W_{3}&=& a^{2}\pa T_{1}+U_{2}T_{1}+T_{2}U_{1}-G^{+}_{2}G^{-}_{1}
        -G^{+}_{1}G^{-}_{2}+a\pa\bvp_{2}T_{1}-a\pa\bvp_{2}T_{2} \CR
     & & -a\pa\bvp_{2}\pa U_{1}-a\bchi_{2}\pa G^{+}_{1}
        -a\pa\bchi_{2}G^{+}_{2}-a\pa^{2}\bvp_{2}U_{2}
        -a^{3}\pa^{3}\bvp_{2},
\label{eq:wcur}
\enn
where
$U_{i}= \chi_{i}\bchi_{i}-a\pa (\bvp_{i}-\vp_{i})$,
$G^{+}_{i}= \chi_{i}\pa\bvp_{i}+a\pa\chi_{i}$,
$G^{-}_{i}= -\bchi_{i}\pa\vp_{i}-a\pa\bchi_{i}$
and
$T_{i}= -\pa\vp_{i}\pa\bvp_{i}-\bchi_{i}\pa\chi_{i}-a\pa^{2}\bvp_{i}$
for $i=1,2$.
$T(z)$ denotes the twisted energy-momentum tensor:
$T(z)=T^{N=2}(z)+\half \pa J(z)$, where $T^{N=2}(z)$ is the
energy-momentum tensor of the $N=2$ model with central charge
$c^{N=2}=6(1+3a^{2})$.
The OPEs of the twisted $N=2$ algebra are given by:
\eqn
T(z)T(w)&=& \ord{2}{2T(w)}+\ordo{\pa T(w)}+\ldots , \CR
T(z)J(w)&=& \ord{3}{-2(1+3a^{2})}+\ord{2}{J(w)}+\ordo{\pa J(w)}+\ldots,\CR
\!\!\!\!\!\!
T(z)G^{+}(w)&=& \ord{2}{G^{+}(w)}
                  +\ordo{\pa G^{+}(w)} +\ldots ,   \ \
T(z)G^{-}(w)=\ord{2}{2G^{-}(w)}
                +\ordo{\pa G^{-}(w)} +\ldots ,   \CR
J(z)J(w)&=& \ord{2}{2(1+3a^{2})} +\ldots , \ \
J(z)G^{\pm}(w)=\ordo{\pm G^{\pm}(w)}+\ldots , \CR
G^{+}(z)G^{-}(w)&=& \ord{3}{2(1+3a^{2})}+\ord{2}{J(w)}+\ordo{T(w)}
+\ldots, \ \
G^{\pm}(z)G^{\pm}(w)=\mbox{regular}.
\label{eq:opts}
\enn
The OPEs between $N=2$ currents and $W$-currents
are
\eqn
J(z)W_{2}(w)&=& \ord{3}{2a^{2}(1+3a^{2})}
               +\ord{2}{(1+3a^{2})J(w)}+\ldots, \CR
J(z)W^{+}(w)&=& \ord{2}{(1+3a^{2})G^{+}(w)}
                     +\ordo{W^{+}(w)}+\ldots, \CR
J(z)W^{-}(w)&=& \ord{2}{(1+2a^{2})G^{-}(w)}
                     +\ordo{-W^{-}(w)}+\ldots, \CR
J(z)W_{3}(w)&=& \ord{4}{4a^{2}(1+3a^{2})}
               +\ord{3}{(1+4a^{2})J(w)}
               +\ord{2}{2W_{2}(w)+(1+2a^{2})T(w)}+\ldots,
\label{eq:opjw}
\enn
\eqn
T(z)W_{2}(w)&=&\ord{4}{-2a^{2}(1+3a^{2})}
               +\ord{3}{-(1+2a^{2})J(w)}
               +\ord{2}{2W_{2}(w)}
               +\ordo{\pa W_{2}(w)}+\ldots, \CR
T(z)W^{+}(w)&=&  \ord{3}{-(1+2a^{2})G^{+}(w)}
                +\ord{2}{2W^{+}(w)}
                +\ordo{\pa W^{+}(w)}+\ldots, \CR
T(z)W^{-}(w)&=& \ord{3}{-G^{-}(w)}
                     +\ord{2}{3 W^{-}(w)}
                     +\ordo{\pa W^{-}(w)}+\ldots, \CR
T(z)W_{3}(w)&=& \ord{3}{-T(w)}
               +\ord{2}{3W_{3}(w)}+\ordo{\pa W_{3}(w)} +\ldots,
\label{eq:optw}
\enn
\eqn
G^{+}(z)W_{2}(w)&=& \ordo{-W^{+}(w)}+\ldots, \ \
G^{+}(z)W^{+}(w)=\mbox{regular}, \CR
G^{+}(z)W^{-}(w)&=& \ord{4}{4a^{2}(1+3a^{2})}
                    +\ord{3}{(1+4a^{2})J(w)}
                    +\ord{2}{2W_{2}(w)}+\ordo{W_{3}(w)}+\ldots, \CR
G^{+}(z)W_{3}(w)&=&
 \ord{3}{(1+4a^{2})G^{+}(w)}+\ord{2}{2W^{+}(w)}+\ldots, \CR
G^{-}(z)W_{2}(w)&=& \ord{2}{a^{2}G^{-}(w)}+\ordo{W^{-}(w)}+\ldots, \ \
G^{-}(z)W^{-}(w)= \mbox{regular}, \CR
G^{-}(z)W^{+}(w)&=& \ord{4}{2a^{2}(1+3a^{2})}
                    +\ord{3}{(1+2a^{2})J(w)}             \CR
                & &    +\ord{2}{-2W_{2}(w)+a^{2}T(w)}
                    +\ordo{W_{3}(w)-\pa W_{2}(w)}+\ldots, \CR
G^{-}(z)W_{3}(w)&=& \ord{3}{G^{-}(w)}+\ord{2}{3W^{-}(w)}
                 +\ordo{\pa W^{-}(w)}+\ldots \ .
\label{eq:opgw}
\enn
Let us discuss the OPEs for $W$-currents.
First we write down the OPEs  between $W_{2}$ and
\{ $W_{2}$, $W^{\pm}$, $W_{3}$ \}:
\eqn
W_{2}(z)W_{2}(w)&=& \ord{4}{1+4a^{2}-a^{4}-12a^{6}}    \CR
 & & +\ord{2}{-2(1+a^{2})W_{2}(w)+(1+2a^{2})[(JJ)+2a^{2}\pa J](w)} \CR
 & & +\ordo{-(1+a^{2})\pa W_{2}(w)+(1+2a^{2})[(\pa JJ)+a^{2}\pa^{2}J](w)}
    +\ldots, \label{eq:op22} \\
W_{2}(z)W^{+}(w)
&=& \ord{3}{(1-4a^{4})G^{+}(w)}
                    +\ord{2}{-(1+a^{2})W^{+}(w)+(1+2a^{2})(J G^{+})(w)}\CR
& & \!\!\!\!\!\!\!\!\!\!\!\!\!\!\!\!\!
  +\ordo{(JW^{+})(w)-(G^{+}W_{2})(w)+(1+2a^{2})(\pa J G^{+})(w)
          -\pa W^{+}(w)}+\ldots, \\
W_{2}(z)W^{-}(w)
&=&  \ord{3}{-(1+2a^{2})G^{-}(w)}  \CR
& &+\ord{2}{-(1+a^{2})W^{-}(w)+(1+2a^{2})(J G^{-})(w)
    +a^{2}(1+2a^{2})\pa G^{-}(w) } \CR
& & +\ordo{\cO^{2,-}_{(1)}(w)}+\ldots,  \\
W_{2}(z)W_{3}(w)&=&
\ord{5}{2(1+4a^{2}-a^{4}-12a^{6})}
+\ord{4}{(1+2a^{2})J(w)} \CR
& & +\ord{3}{-2(1+a^{2})W_{2}(w)+(1+2a^{2})[-a^{2}T+(JJ)+3a^{2}\pa J ](w)}
\CR
& & +\ord{2}{\cO^{2,3}_{(2)}(w)}+\ordo{\cO^{2,3}_{(1)}(w)}+\ldots,
\enn
where
\eqn
\cO^{2,-}_{(1)}(w)&=&-(JW^{-})(w)+(G^{-}W_{2})(w)
    -(1+a^{2})\pa W^{-}(w)\CR
& & +(1+a^{2})(\pa J G^{-})(w)+a^{2}(J\pa G^{-})(w)
    +a^{2}(\half+a^{2})\pa^{2}G^{-}(w), \CR
\cO^{2,3}_{(2)}(w)
&=&  -W_{3}(w)+(JW_{2})(w)-\pa W_{2}(w)-a^{2}(G^{+}G^{-})(w)
    +(1+a^{2})(JT)(w) \CR
& &  +a^{2}\pa T(w) +(1+2a^{2})[(\pa JJ)+{3a^{2}\over 2}\pa^{2}J](w), \CR
\cO^{2,3}_{(1)}(w)
&=& (G^{+}W^{-})(w)+(G^{-}W^{+})(w)+a^{2}(\pa G^{-} G^{+})(w)
   +(\pa JW_{2})(w)-\pa W_{3}(w) \CR
& & +(1+a^{2})(\pa J T)(w)+(a^{2}+\half)(\pa^{2}JJ)(w)
    +a^{2}(a^{2}+{2\over3})\pa^{3}J(w).
\label{eq:oc23}
\enn
Here a normal ordered product $(AB)(z)$ for two generators $A(z)$ and
$B(z)$ are defined as
\eq
(AB)(z)=\int {d w\over 2\pi\I}{A(w)B(z)\over w-z}.
\en
Second the OPEs  among fermionic $W$-currents
$W^{\pm}$ are given by
\eqn
W^{+}(z)W^{+}(w)&=&\ordo{-(1+2a^{2})(G^{+}\pa G^{+})(w)+2(G^{+}W^{+})(w)}
                 +\ldots, \CR
W^{-}(z)W^{-}(w)&=&\ordo{-(G^{-}\pa G^{-})(w)-2(G^{-}W^{-})(w)}
                 +\ldots, \CR
W^{+}(z)W^{-}(w)
&=& \ord{5}{2(1+4a^{2}-a^{4}-12a^{6})}
    +\ord{4}{(1+2a^{2})J(w)}  \CR
& & \!\!\!\!\!\!\!\!
 +\ord{3}{(1+a^{2}-2a^{4}) T(w)-2(1+a^{2})W_{2}(w)
            +(1+2a^{2})[(JJ)+3a^{2}\pa J](w)} \CR
& & \!\!\!\!\!\!\!\!
+\ord{2}{\cO^{+,-}_{(2)}(w)}+\ordo{\cO^{+,-}_{(1)}(w)}+\ldots,
\label{eq:oppm}
\enn
where
\eqn
\cO^{+,-}_{(2)}(w)
&=&a^{2}W_{3}(w)-\pa W_{2}(w)+(JW_{2})(w)+(1+a^{2})(G^{+}G^{-})(w) \CR
& & -a^{2}(JT)(w)-2a^{4}\pa T(w)
    +(1+2a^{2})[(\pa J J)+{3a^{2}\over 2}\pa^{2}J](w), \CR
\cO^{+,-}_{(1)}(w)
&=& (JW_{3})(w)-(TW_{2})(w)+(\pa J W_{2})(w)+a^{2}\pa W_{3}(w)
    -a^{2}(\pa T J)(w) \CR
& & \!\!\!\!\!\!\!\! \!\!\!\!\!\!\!\!  \!\!\!\!\!\!\!\!
  +(1+a^{2})(\pa G^{+} G^{-})(w)
  -a^{4}\pa^{2}T(w)+(a^{2}+\half)(\pa^{2}J J)
  +a^{2}(a^{2}+{2\over 3})\pa^{3}J(w).
\label{eq:ocpm}
\enn
Third the OPEs between $W^{\pm}$ and $W_{3}$ are
\eqn
W^{+}(z)W_{3}(w)&=& \ord{4}{(1+2a^{2})G^{+}(w)} \CR
& & +\ord{3}{-2(1+a^{2})W^{+}(w)
       +(1+2a^{2})[ (J G^{+})+(G^{+}J)+3a^{2}\pa G^{+}](w)} \CR
& &+\ord{2}{\cO^{+,3}_{(2)}(w)}+\ordo{\cO^{+,3}_{(1)}(w)}+\ldots,\CR
W^{-}(z)W_{3}(w)&=&
\ord{4}{4(1+2a^{2})G^{-}(w)}+\ord{3}{2(1+2a^{2})\pa G^{-}(w)}
+\ord{2}{\cO^{-,3}_{(2)}(w)}+\ordo{\cO^{-,3}_{(1)}(w)}+\ldots, \CR
\label{eq:oppm3}
\enn
where
\eqn
\cO^{+,3}_{(2)}(w)
&=&-\pa W^{+}(w)+(J W^{+})(w) +(G^{+}W_{2})(w)+(TG^{+})(w) \CR
& & +(1+2a^{2})[ (\pa J G^{+})+(\pa G^{+} J)](w)
    +(-\half+{3a^{2}\over 2}+3a^{4})\pa^{2}G^{+}(w), \CR
\cO^{+,3}_{(1)}(w)
&=& [(G^{+}W_{3})-(TW^{+})+(\pa J W^{+})+(\pa G^{+}W_{2})](w)
    +(1+a^{2})(\pa G^{+} T)(w) \CR
& & \!\!\!\!\!\!\!\!\!\!\!\!\!\!\!\!\!\!\!\!\!\!\!
    -a^{2}(\pa T G^{+})(w)
    +(\half +a^{2})[(\pa^{2}J G^{+})+(\pa^{2}G^{+}J)](w)
    +(a^{4}+{2a^{2}\over 3})\pa^{3}G^{+}(w),
\label{eq:ocp3}
\enn
and
\eqn
\cO^{-,3}_{(2)}(w)
&=& 2(JW^{-})(w)+(G^{-}T)(w)+2a^{2}\pa W^{-}(w) \CR
& & +(\pa G^{-}J)(w)-(\pa J G^{-})(w)+2a^{2}\pa^{2}G^{-}(w) , \CR
\cO^{-,3}_{(1)}(w)
&=& -(G^{-}W_{3})(w)+(TW^{-})(w)+(J\pa W^{-})(w)+(\pa J W^{-})(w)
    +(\pa G^{-}T)(w) \CR
& & +a^{2}\pa W^{-}(w)+\half (\pa^{2}G^{-}J)(w)-\half (\pa^{2}J G^{-})(w)
    +{2a^{2}\over 3}\pa^{3}G^{-}(w).
\label{eq:ocm3}
\enn
Finally the OPE between $W_{3}$ and itself is
found to be
\eq
W_{3}(z)W_{3}(w)
=\ord{4}{4(1+2a^{2})T(w)}
+\ord{3}{2(1+2a^{2})\pa T(w)}
+\ord{2}{\cO^{3,3}_{(2)}(w)}+\ordo{\cO^{3,3}_{(1)}(w)}+\ldots,
\label{eq:op33}
\en
where
\eqn
\cO^{3,3}_{(2)}(w)
&=& 2(JW_{3})(w)-2(G^{+}W^{-})(w)+(TT)(w)-(\pa J T)(w)+(\pa T J)(w) \CR
& & +2a^{2}\pa W_{3}(w)+2a^{2}\pa^{2} T(w)
   +[(\pa G^{+}G^{-})+(\pa G^{-}G^{+})](w), \CR
\cO^{3,3}_{(1)}(w)&=& \half \pa\cO^{3,3}_{(2)}(w)
-{1\over6}(1+2a^{2})\pa^{3}T(w).
\label{eq:oc33}
\enn
The equations (\ref{eq:opts})-(\ref{eq:opgw}) and
(\ref{eq:op22})-(\ref{eq:oc33})
complete the operator product expansions of the $N=2$ super $W_{3}$
algebra.
It is easy to see that the present definitions of the $W$-currents
(\ref{eq:wcur}) give the Poisson algebra structure
with quadratic relations in the classical limit of
$a^{2}\rightarrow -\infty$ \cite{Ito2}.

Note that in the present definitions (\ref{eq:wcu}), the $W$-currents
are quasi-primary fields with respect to the energy-momentum tensor
$T^{N=2}(z)$.
One may add suitable differential polynomials of the currents to
these $W$-currents in order to get a $N=2$ supermultiplet
$\{ \tW_{2}, \tW^{\pm}, \tW_{3} \}$ \cite{NeYa,AM2}:
\eqn
\tW_{2}&=&W_{2}-{a^{2}\over 2}\pa J-{3+8a^{2} \over 2(5+18a^{2})}(JJ)
        +{(1-2a^{2})(1+3a^{2})\over 5+18a^{2}}T^{N=2}, \CR
\tW^{+}&=&W^{+}+{(1-2a^{2})(1+3a^{2})\over 5+18a^{2}}\pa G^{+}
          -{3+8a^{2} \over 5+18a^{2}}(JG^{+}), \CR
\tW^{-}&=&W^{-}-{1+6a^{2}+12a^{4}\over 5+18a^{2}}\pa G^{-}
          -{3+8a^{2} \over 5+18a^{2}} (JG^{-}), \CR
\tW_{3}&=& W_{3}-\half\pa W_{2}
          +{3+8a^{2} \over 5+18a^{2}}
           [ (G^{+}G^{-})-(JT) +{1\over 4}\pa (JJ)] \CR
       & & -{3+13a^{2}+18a^{4}\over 2(5+18a^{2})}\pa T^{N=2}
           -{(2+3a^{2})(1+2a^{2})\over 4(5+18a^{2})}\pa^{2} J.
\label{eq:mw}
\enn
{}From (\ref{eq:mw}) and the OPEs for the $W$-currents, we may compute
the operator product expansions
for the $N=2$ super-multiplet $\{ \tW_{2}, \tW^{\pm}, \tW_{3} \}$.
After some calculations, one finds that these operator
product expansions coincide with the results of the OPE method
\cite{Ro}.
For example, one gets
\eq
\tW_{2}(z)\tW_{2}(w)=\ord{4}{{c^{N=2}\B^{2}\over 2}}
+\ord{2}{\widetilde{\cO}^{2,2}_{(2)}(w)}
+\ordo{\half\pa\widetilde{\cO}^{2,2}_{(2)}(w)}+\ldots ,
\en
where $\B=\sqrt{{(1-2a^{2})(2+3a^{2})(1+4a^{2})\over 3(5+18a^{2})}}$
and
\eq
\widetilde{\cO}^{2,2}_{(2)}(w)
=-{6(1+2a^{2})(1+5a^{2})\over (5+18a^{2})}\tW_{2}(w)
 +{2c^{N=2}\B^{2}\over 5+18a^{2}}[T^{N=2}-{1\over 4(1+3a^{2})}(JJ)](w).
\en
The relation between the present definitions and those
in ref. \cite{Ro} is given by formulae
$(J,G^{+}, G^{-},T^{N=2})=(J,\sqrt{2}G,\sqrt{2}\bar{G},T)$ and
$(\tW_{2},\tW^{+},\tW^{-},\tW_{3})=-\B (V,\sqrt{2}U,\sqrt{2}\bar{U},W)$,
where $(J,G,\bar{G},T)$ and $(V,U,\bar{U},W)$ are the $N=2$ primary
$W$-currents in ref. \cite{Ro}.

It is known that the $A(n|n-1)$ Miura transformations give the
Feigin-Fuchs representation of the $N=2$ $CP_{n}$ super coset
models\cite{KaSu} for $a^{2}=-1/(k+n+1)$ \cite{Ito1,NeYa}.
In these $CP_{n}$ coset models there is a kind of duality relation
such that \{ $CP_{n}$ models at level $k=1$ \}
$\equiv$ \{ $CP_{1}$ models at level $k=n$\} \cite{KaSu}.
In the case of $n=2$, we get $a^{2}=-1/4$.
It easy to see that the $W$-currents $\tW_{2},\tW^{+},\tW^{-},\tW_{3}$
have no central extensions and become zero-norm
fields at this duality point\footnote{
A similar phenomena has been observed in the case of
$Z_{k}$ parafermions\cite{NaQu}.
The author would like to thank Wolfgang Lerche for pointing out this fact.
}.
Therefore we expect that these $W$-currents decouples from the
representations of $N=2$ super-$W_{3}$ algebra\cite{BeLeNeWa}.
But in order to check the decoupling we need to investigate the
structure of the Verma modules more carefully, which is
left for further investigation.

Although the present free field realization is obtained from the
supersymmetric Miura transformation (\ref{eq:miura}), the algebra does
not depend on the particular free field realization of
$U_{1}$, $G^{\pm}_{1}$ and $T_{1}$.
This suggests the existence of other types of the free field realization
of $N=2$ super-$W_{3}$ algebra, which would be useful for
studying the quantum topological $W_{3}$-gravity theory.

The author would like to thank Wolfgang Lerche and Xiang Shen for
useful discussions and the CERN Theory Division for its hospitality,
where a part of this work was done.

\end{document}